\documentclass[pra,twocolumn,superscriptaddress]{revtex4-1}

\usepackage{amsmath,color,amsfonts,graphicx,bm}

\newtheorem{conj}{Conjecture}

\newcommand{\ket}[1]{\vert#1\rangle}
\newcommand{\bra}[1]{\langle#1\vert}
\newcommand{\mc}[1]{\mathcal{#1}}
\newcommand{\tr}{\mathrm{tr}}
\newcommand{\even}{\mathrm{even}}
\newcommand{\odd}{\mathrm{odd}}
\newcommand{\lrket}[1]{\left\vert#1\right\rangle}

\begin{document}

\title{Topological order in PEPS:
 Transfer operator and boundary Hamiltonians}

\author{Norbert Schuch}
\affiliation{Institut f\"ur Quanteninformation, 
    RWTH Aachen, 52056 Aachen, Germany}
\affiliation{Institute for Quantum Information, California Institute of
Technology, MC 305-16, Pasadena CA 91125, U.S.A.}
\author{Didier Poilblanc}
\affiliation{Laboratoire de Physique Th\'eorique, C.N.R.S.\ 
    and Universit\'e de Toulouse, 31062 Toulouse, France}
\author{J.~Ignacio \surname{Cirac}}
\affiliation{Max-Planck-Institut f\"ur Quantenoptik,
Hans-Kopfermann-Str.\ 1, D-85748 Garching, Germany}
\author{David P\'erez-Garc\'ia}
\affiliation{Dpto.\ Analisis Matematico and IMI, 
    Universidad Complutense de Madrid, 
    E-28040 Madrid, Spain}

\begin{abstract}
We study the structure of topological phases and their boundaries in the
Projected Entangled Pair States (PEPS) formalism.  We show how topological
order in a system can be identified from the structure of the PEPS
transfer operator, and subsequently use these findings to analyze the
structure of the \emph{boundary Hamiltonian}, acting on the bond
variables, which reflects the entanglement properties of the system. We
find that in a topological phase, the boundary Hamiltonian consists of two
parts: A universal non-local part which encodes the nature of the
topological phase, and a non-universal part which is local and inherits
the symmetries of the topological model, which helps to infer the
structure of the boundary Hamiltonian and thus possibly of the physical
edge modes.  \end{abstract}

\maketitle


The study of strongly correlated quantum systems is of central interest in
modern condensed matter physics due to the exciting properties exhibited
by those systems, in particular unconventional phases with topological
order. In order to identify topological order in such systems, topological
entropies~\cite{kitaev:topological-entropy,levin:topological-entropy}
have been applied successfully.  To obtain
more information than contained in
the entropy, the entanglement spectrum (ES) -- i.e., the
spectrum of the reduced density operator of a region -- has been
studied, and it has been realized that for certain systems, the low-energy
part of the ES resembles the spectrum of the thermal state of a
one-dimensional (1D) local \emph{boundary Hamiltonian} which can be
associated to the boundary of the region studied, and which seems to be
related 
to the physical edge modes of the
model~\cite{li:es-qhe-sphere,poilblanc:es-heisenberg-ladder,%
laeuchli:es-spinladder,schliemann:es-hall-bilayers}. While this relation
between bulk ES, boundary Hamiltonian, and edge
excitations can be made rigorous in some
cases~\cite{fidkowski:freeferm-bulk-boundary,qi:bulk-boundary-duality}, in
most cases the Hamiltonian is \emph{a posteriori} inferred
from the structure of the ES, and a general connection
between ES and boundary still needs to be made.

In~\cite{cirac:peps-boundaries}, we made progress in that direction
by proving a rigorous connection between ES and
boundary using the framework of Projected Entangled Pair States
(PEPS)~\cite{verstraete:2D-dmrg}, which  form the appropriate description
of ground states of gapped local Hamiltonians both in conventional and
topological phases~\cite{hastings:mps-entropy-ent}.  This allowed us to
derive a one-dimensional boundary Hamiltonian, which we found to be
local in trivial phases (without symmetry breaking or topological order).
Following the Li-Haldane conjecture about the relation of boundary
Hamiltonian and edge physics~\cite{li:es-qhe-sphere}, this allows for
conclusions about the structure of a system's edge excitations.  On the
other hand, for both symmetry-broken and topological phases we found a
highly non-local boundary Hamiltonian, making it impossible to infer
something about the actual edge physics.  Yet, since this boundary
Hamiltonian acts on the virtual bond variables, its local and non-local
character are not necessarily reflected in physical space.

In this paper, we establish a framework for studying the boundary
Hamiltonian of topologically ordered phases in the framework of PEPS.  We
start by showing how topological order is reflected in the properties 
of the transfer operator, which in turn enables us to decompose the
boundary Hamiltonian of topological models into two parts. The
\emph{universal part} couples to non-local (topological) degrees of
freedom and determines the phase of the system, but is independent on
microscopic details.  The \emph{non-universal} part is local (thereby
generalizing what happens for trivial phases), depends on microscopic
details, but vanishes under RG flows; moreover, it commutes with the
symmetries which originate from the universal part. Therefore, the
non-universal part can help to infer the nature of the edge physics of the
model.

\begin{figure}[t]
\includegraphics[width=\columnwidth]{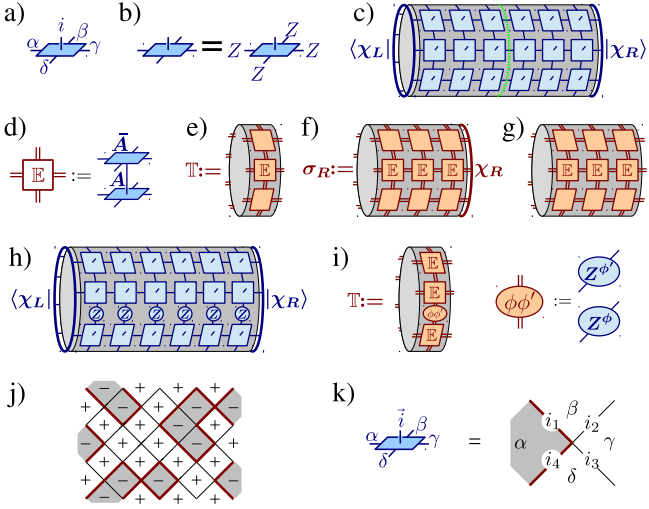}
\caption{\label{fig:peps}
Tensor networks for entanglement spectra and for topological models; see
text for details.} 
\end{figure}

Let us first introduce PEPS and explain how to use them to derive 
boundary theories. For clarity, we restrict to square
lattices on a cylinder (with length $N_h$ and circumference $N_v$). A
(translational invariant) PEPS $\ket{\psi}=\sum c_{i_1\dots
i_N}\ket{i_1,\dots,i_N}$ is described by a five-index tensor
$A^i_{\alpha\beta\gamma\delta}$ (Fig.~\ref{fig:peps}a, with $i$ the
\emph{physical} and $\alpha,\beta,\gamma,\delta$ the \emph{virtual}
indices), such that the coefficient $c_{i_1\dots i_N}$ is obtained by
arranging tensors $A^{i_1},\dots,A^{i_N}$ on the cylinder and contracting
each virtual index with the corresponding index of the adjancent tensors,
while putting boundary conditions $\ket{\chi_L}$, $\ket{\chi_R}$ at the
open virtual indices at the two ends (Fig.~\ref{fig:peps}c). 
PEPS naturally appear as ground states of local \emph{parent
Hamiltonians}~\cite{perez-garcia:parent-ham-2d,schuch:peps-sym}; the
boundary conditions $\ket{\chi}$ can be incorporated in the Hamiltonian by
making the virtual indices at the boundary physical and including them in the
parent Hamiltonian, and additionally acting with a frustration-free Hamiltonian
term with ground space $\ket{\chi}$ on them. (In particular, if $\ket\chi$
is a Matrix Product State, this Hamiltonian is local.)

As proven in~\cite{cirac:peps-boundaries} (see also
Appendix~A), for any PEPS the ES
of a half-cylinder (green cut in Fig.~\ref{fig:peps}c) is equal to the
spectrum of 
\begin{equation}
\label{eq:sym-bnd}
\sigma\propto\sqrt{\sigma_L^*}\sigma_R\sqrt{\sigma_L^*}\ ,
\end{equation}
where $\sigma_R$ is the state obtained
at the \emph{virtual} indices of the right half-cylinder by contracting
the physical indices with the adjoint, cf.  Fig.~\ref{fig:peps}d,f (with
$\chi_R=\ket{\chi_R}\bra{\chi_R}$, and correspondingly for $\sigma_L$);
for $N_h\gg 1$, this is just the eigenvector corresponding to the
largest eigenvalue of the \emph{transfer operator} $\mathbb T$,
Fig.~\ref{fig:peps}d,e.  From there, one can construct a
\emph{boundary Hamiltonian} $H=-\log\sigma$ which acts on the virtual
degrees of freedom at the boundary, and which exactly reproduces the
ES.  As demonstrated in~\cite{cirac:peps-boundaries},
this Hamiltonian is local if the system is in a trivial phase, and becomes
non-local in symmetry broken or topological phases.  In this work, we
revisit the structure of the boundary Hamiltonian for topological
phases: There, the transfer operator exhibits symmetries and
degeneracies, giving rise to a non-unique fixed point. As we will show, by
properly interpreting the structure of the transfer operator and
identifying the physically relevant fixed points, the locality of the
boundary Hamiltonian can in part be recovered also for topological phases.

Topological order in PEPS is accompanied by
a virtual symmetry of the tensor $A$, such as the invariance under the
representation of a (finite) symmetry group,
Fig.~\ref{fig:peps}b~\cite{schuch:peps-sym} (more general symmetries are
given, e.g., by Hopf algebras~\cite{buerschaper:topo-TNs} or tensor
categories~\cite{levin:stringnets}).  For simplicity, we focus on
$\mathbb Z_2$ symmetry, i.e., $A$ is invariant under $Z^{\otimes 4}$, for
some unitary representation $\{\openone,Z\}$ of $\mathbb Z_2$, but our
findings easily generalize to any finite group. In that case, the four
possible ground states are distinguished by \emph{(i)} whether
$\ket{\chi_L}$ and $\ket{\chi_R}$ are in the $\pm1$ eigenspace of
$Z^{\otimes N_v}$ (i.e., have an even/odd parity of $\ket1$'s, denoted
$p=e,o$), and \emph{(ii)} by the possibility of having a string of $Z$'s
along the cylinder, Fig.~\ref{fig:peps}h~\cite{schuch:peps-sym}.  It
is convenient to picture the $Z$ string as coupled to a flux
$\phi\in\{0,\pi\}$ threading the cylinder.

The symmetry Fig.~\ref{fig:peps}b of $A$ induces the same symmetry
independently in the ket and bra layer of $\mathbb E$, 
Fig.~\ref{fig:peps}d, and subsequently in
$\mathbb T$ (Fig.~\ref{fig:peps}e), i.e., $[\mathbb
T,Z^{\otimes N_v}\otimes \openone^{\otimes N_v}]= [\mathbb
T,\openone^{\otimes N_v}\otimes Z^{\otimes N_v}]= 0$, where the tensor
product is w.r.t.\ ket and bra layer; this is, $\mathbb T$ has four blocks
corresponding to the $Z^{\otimes N_v}$ eigenvalue (i.e., parity) for both
the ket and the bra layer.  If we include the $Z$ string coupled to the
flux, Fig.~\ref{fig:peps}h, we find that the overall transfer
operator consists of four such transfer operators $\mathbb T_\phi^{\phi'}$
(Fig.~\ref{fig:peps}i), each corresponding to a flux $\phi$ for the
ket and $\phi'$ for the bra layer, respectively.  Overall, it follows that
the transfer operator is block-diagonal with $16$ blocks
$\mathbb T_{p\phi}^{p'\phi'}$, each
corresponding to one of the $16$ blocks $\rho_{p\phi}^{p'\phi'}$ 
($\rho\equiv\sigma_L,\sigma_R$) of the
fixed point density operator, with parity $p$ ($p'$) and flux $\phi$
($\phi'$) on the ket (bra) layer.

As an example, let us consider Kitaev's Toric Code
(TC)~\cite{kitaev:toriccode}. Locally, it is a uniform superposition of
all closed loops on a lattice, which can be described by assigning dual
variables $\ket{\pm}$ (colors) to the plaquettes, with loops
whereever the dual variable changes (i.e., loops are boundaries of
colored regions), Fig.~\ref{fig:peps}j. The PEPS is then obtained by
blocking the marked region and assigning the bonds to the plaquette
variables, Fig.~\ref{fig:peps}k. The $Z^{\otimes 4}$ symmetry of the tensor
reflects the fact that inverting the entire coloring does not change
the state.  A $Z$ string along the cylinder (Fig.~\ref{fig:peps}h) flips
the coloring, which leads to an odd number of horizontal strings; while an
even (odd) $Z^{\otimes N_v}$ parity in $\ket{\chi_L}$ and $\ket{\chi_R}$
gives a state with a plus (minus) superposition of an even and odd number
of loops around the cylinder.

For the TC, $\mathbb E=(\openone^{\otimes
4}+Z^{\otimes 4})$, and thus 
$\mathbb T_\phi^{\phi}=
    \tfrac12(\openone^{\otimes N_v}\otimes \openone^{\otimes N_v}+ 
    Z^{\otimes N_v}\otimes Z^{\otimes N_v})
=P_{e}\otimes P_{e}+
P_o\otimes P_o$,
with 
$P_{e/o}=\tfrac12(\openone^{\otimes N_v}\pm Z^{\otimes N_v})$
the projectors onto the even/odd parity subspace at the boundary, while
for $\phi'\ne\phi$, $\mathbb T_\phi^{\phi'}=0$.
This is, $\mathbb T$ has
four degenerate fixed points, corresponding to the four ``diagonal''
blocks $\rho_{p\phi}^{p\phi}$ of the boundary. The four blocks correspond to
the four ground states, and, as we will see, their degeneracy
is essential for the system to be topologically ordered.

To better understand how the structure of the transfer operator reflects
the order of the system, we add string tension to the TC,
i.e., weigh every configuration with $\lambda^{\ell}$, where $\ell$ is the
total length of all loops; this can be achieved by locally modifying the
tensors (keeping the $Z^{\otimes 4}$ symmetry), and translates to
a magnetic field in the
Hamiltonian~\cite{castelnovo:tc-tension-topoentropy,trebst:tcode-phase-transition}.
This model
exhibits a topological phase transition at
$\lambda_\mathrm{crit}=1/\sqrt{1+\sqrt 2}\approx 0.644$ (with $\lambda=1$
the TC and $\lambda=0$ the
vacuum)~\cite{castelnovo:tc-tension-topoentropy}.

\begin{figure}
\includegraphics[width=\columnwidth]{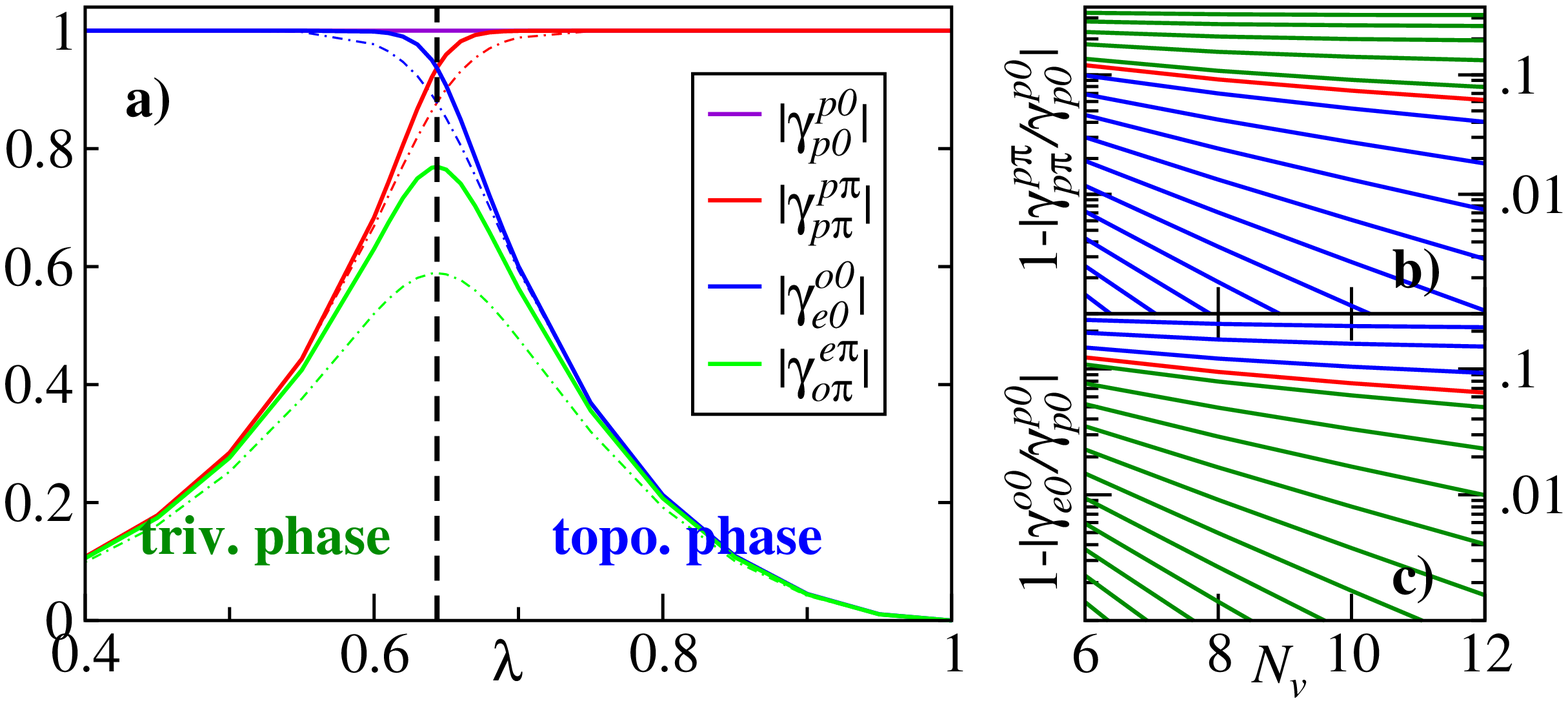}
\caption{
\label{fig:tc-topspec}
\textbf{a)} Maximal eigenvalues $\gamma^{p'\phi'}_{p\phi}$ for the TC with
string tension (relative to $\gamma_{e0}^{e0}$) for
$N_v=12$ (solid lines) and $N_v=6$ (dash-dotted lines).
\textbf{b,c)} Splitting
$1-|\gamma_{p\pi}^{p\pi}/\gamma_{p0}^{p0}|$ (b) and 
$1-|\gamma_{e0}^{o0}/\gamma_{p0}^{p0}|$ (c) for
$\lambda=0.65,0.66,\dots$ (blue, topological phase), and
$\lambda=0.64,0.63,\dots$ (green, trivial phase). The red line is
$\lambda = 0.644\approx \lambda_\mathrm{crit}$.
The calculations (as well as the ones in Figs.~\ref{fig:ham-locality} and
\ref{fig:rvb}) have been carried out using exact column-wise contraction
(cf.~\cite{cirac:peps-boundaries,schuch:rvb-kagome,poilblanc:rvb-boundaries})
and are thus exact.
}
\end{figure}

Fig.~\ref{fig:tc-topspec}a shows the modulus of the largest eigenvalue
$\gamma_{p\phi}^{p'\phi'}$ for each block $\mathbb T_{p\phi}^{p'\phi'}$ of
the transfer operator. We first focus on the topological phase:
We find that the four ``diagonal'' blocks $\mathbb
T^{p\phi}_{p\phi}$ are essentially degenerate (the splitting vanishes
exponentially with $N_v$, see Fig.~\ref{fig:tc-topspec}b), which ensures
that there are four stable ground states.  At the same time, the
off-diagonal
blocks are strictly smaller than the diagonal blocks, which ensures that
the four states are linearly independent in the thermodynamic limit 
(see later).  In addition, we find that the diagonal blocks $\mathbb
T_{p\phi}^{p\phi}$ are gapped (not shown), which ensures that each block
has a unique fixed point $\rho_{p\phi}^{p\phi}$.

Altogether, we find that the fixed point of the transfer operator is a
direct sum of the $\rho_{p\phi}^{p\phi}$ (i.e., block-diagonal), with
weights determined by
the boundary condition $\ket{\chi}$. Symmetrizing,
Eq.~(\ref{eq:sym-bnd}), preserves this block structure, and we find that
the density operator $\sigma_\mathrm{topo}$ which reproduces the
ES is of the form 
\begin{equation}
\label{eq:sigma-topo}
\sigma_\mathrm{topo} = w_{e0}^{e0}\, \sigma_{e0}^{e0} \;\oplus\;
	     w_{o0}^{o0}\, \sigma_{o0}^{o0} \;\oplus\;
	     w_{e\pi}^{e\pi}\, \sigma_{e\pi}^{e\pi} \;\oplus\;
	     w_{o\pi}^{o\pi}\, \sigma_{o\pi}^{o\pi} \ ,
\end{equation}
where the weights $w_{p\phi}^{p\phi}\ge0$ can be adjusted arbitrarily by
appropriate boundary conditions. We can now define a Hamiltonian
$H=-\log\sigma$ which reproduces the ES. $H$ commutes
with both $Z$ parity and flux, this is, there are 
$H_\phi$, $\phi=0,\pi$, satisfying $[H_\phi,Z^{\otimes N_v}]=0$,
i.e., the $H_\phi$ obey a superselection rule inherited from the topological
symmetry.

Hamiltonians with different weights $w_{p\phi}^{p\phi}$ ($H_\phi)$ and
$q_{p\phi}^{p\phi}$ ($H'_\phi$) are related via
$H'_\phi = H_\phi -
\log(q_{e\phi}^{e\phi}/w_{e\phi}^{e\phi})\,P_e
\oplus 
\log(q_{o\phi}^{p\phi}/w_{o\phi}^{o\phi})\,P_o
$.
This implies
two things: First, the boundary
Hamiltonians obtained for different boundary conditions differ just by a
\emph{universal} contribution which only depends on the underlying
symmetry but not on microscopic details. Second, since
$P_{e/o}=\tfrac12(\openone^{\otimes N_v}\pm Z^{\otimes N_v})$, the
boundary Hamiltonian will generally be highly non-local, 
and, if at all, 
will only be local for a specific choice $w_{p\phi}^{p\phi}$.

\begin{figure}
\includegraphics[width=\columnwidth]{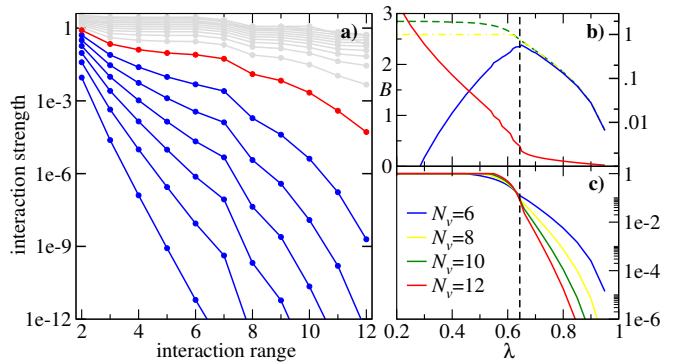}
\caption{\label{fig:ham-locality}
\textbf{a}) Interaction range of $H_0$ 
for the TC with string tension
($N_v=12$), obtained using Eq.~(\ref{eq:sigma-topo}) with equal weights, which is
the correct interpretation for the topological sector.
Interactions in the topological phase (blue,
$\lambda=0.7,0.75,\dots$) decay exponentially.  In the trivial phase
(grey, $\lambda=0.60,0.55,\dots$), the interpretation is no longer valid,
seemingly leading to a non-local Hamiltonian.  (The red line is
$\lambda_\mathrm{crit}$.)
\textbf{b)}~By changing to the interpretation of the transfer operator
which is valid in the trivial phase (see text), we obtain a $H_0$
which is local in the trivial phase. The blue line shows
$\|\sigma_\mathrm{triv}-\mu(B)\|_1/{N_v}$ for $N_v=12$ (right scale),
cf.~text, and the red line the corresponding $B$ (left scale).  The
comparison with $\|\sigma_\mathrm{triv}-\mu(B=0)\|_1/{N_v}$ (green) and
$\|\sigma_\mathrm{topo}-\mu(B)\|_1/{N_v}$ [yellow, 
cf.~Eq.~(\ref{eq:sigma-topo})],
shows that the boundary Hamiltonian is well
approximated by $B\sum X_i$ in the trivial phase, while the decay in the
topological phase is due to the high temperature.  \textbf{c)} Comparison
of $H_0$ and $H_\pi$. The plot shows $\|H_0-\mathcal
F(H_\pi)\|_\mathrm{op}/ \|H_0-H_\pi\|_\mathrm{op}$, where $\mathcal F$
flips the sign of all terms which change the parity across
the boundary (see text); in the topological phase, the difference
convergences to $0$ exponentially in $N_v$.  } \end{figure}

As discussed in Appendix~B, the only choice for which we can expect a
local Hamiltonian is $w_{e\phi}^{e\phi}=w_{o\phi}^{o\phi}$.  The result
for the TC with string tension is shown in
Fig.~\ref{fig:ham-locality}, and we find indeed that the terms in $H_\phi$
decay exponentially with distance, i.e., $H_\phi$ is local (see
Appendix~C).
 Note that by combining the locality of $H_\phi$
with the symmetry $[H_\phi,Z^{\otimes N_v}]=0$, we can already infer that
$H_\phi$ must be well approximated by a parity preserving nearest neighbor
Hamiltonian; in the language of creation/annihilation operators, this
amounts to hopping, pairing, repulsion, and on-site potential terms.
Closer analysis yields that $H_\phi$ is very well approximated by an
Ising Hamiltonian $\sum
X_iX_{i+1}$, with strongly supressed longer-range Ising couplings, and
even more strongly supressed many-body terms.  One would naturally expect
that $H_0$ and $H_\pi$ only differ by a phase $e^{i\pi}=-1$ for
terms which change the parity across the boundary,
 which is indeed what we observe
(Fig.~\ref{fig:ham-locality}c).

Figure~\ref{fig:ham-locality} also shows that $H_\phi$ becomes long-ranged
at the phase transition and stays so in the trivial phase, seemingly
contradicting earlier findings~\cite{cirac:peps-boundaries}
where the Hamiltonian in the trivial phase was local.  However, the
derivation of $H_\phi$ was based on the structure of the transfer
operator, which changes radically in the trivial phase 
(Fig.~\ref{fig:tc-topspec}): First, eigenvalues corresponding to $\phi=\pi$,
$\gamma^{p\pi}_{p\pi}$, become strictly smaller than one.  This implies
that the norm of states in the $\phi=\pi$ sector vanishes
exponentially in $N_h$, and thus, states in that sector are unstable: 
Any random symmetry-preserving
perturbation of $A$ (and thus of the parent Hamiltonian) will yield an
admixture of the $\phi=0$ sector at $N_v$th order, and thus, for an
appropriate ratio $N_v/N_h$, the $\phi=\pi$ sector vanishes in the
thermodynamic limit.  It remains to see what
happens to the two states in the $\phi=0$ sector.  There, 
$|\gamma_{e0}^{o0}|\rightarrow\gamma_{e0}^{e0}=\gamma_{o0}^{o0}$ (see
Fig.~\ref{fig:tc-topspec}c), which implies that the two states in
this sector become equal in the thermodynamic limit, since their overlap
is given by $\tr[\mathbb T_{e0}^{o0}]/\sqrt{\tr[\mathbb
T_{e0}^{e0}]\tr[\mathbb T_{o0}^{o0}]}\rightarrow (\gamma_{e0}^{o0}/
\sqrt{\gamma_{e0}^{e0}\gamma_{o0}^{o0}})^{N_h}\rightarrow 1$.
Thus, studying the transfer operator reveals that the
system in fact has only one ground state. 

In accordance with the changed structure of the transfer operator in the
trivial phase, the boundary state $\sigma_\mathrm{triv}$ can be any state
$\sigma_\mathrm{triv} = \sum_{p,p'}
w_{p\,0}^{p'0}\sigma_{p\,0}^{p'0}\ge0$, which all have the same spectrum
but yield different Hamiltonians.   Choosing
$w_{e0}^{e0}=w_{o0}^{o0}=\tfrac12$ and extremal $w_{e0}^{o0}=w_{o0}^{e0}$
(such that $\sigma_\mathrm{triv}$ becomes singular), we find that
$\sigma_\mathrm{triv}$ is well approximated by $\mu(\beta)=\exp[-B\sum
X_i]/Z$, see Fig.~\ref{fig:ham-locality}b.

Let us briefly summarize our findings: We have found that the virtual
symmetry of topological PEPS (Fig.~\ref{fig:peps}b) induces a 
block-diagonal structure of the transfer operator; topological order is
witnessed by the degeneracy of the diagonal blocks. 
We can then construct boundary Hamiltonians
$H'_\phi=\beta_\mathrm{topo}H_\mathrm{topo}+H_\phi$, with a
\emph{universal part} $H_\mathrm{topo} = Z^{\otimes N_v}$ which only
depends on the symmetry (which is universal), and a $\beta_\mathrm{topo}$
which depends on the boundary conditions. The \emph{non-universal part}
$H_\phi$ is local (i.e., vanishes under RG) and thus represents the
short-range physics of the system, and it is independent of boundary
conditions.  A
phase transition is accompanied by a diverging interaction length of
$H_\phi$.  $H_\phi$ inherits the PEPS symmetry, $[H_\phi,Z^{\otimes
N_v}]=0$, which -- together with the locality of $H_\phi$ -- allows to
infer much of its structure, and it couples to the flux in a natural way.
Note that the symmetry also constrains the structure of the physical edge
modes: The space of zero-energy excitations is spanned by putting
arbitary boundary conditions $\ket{b}$ with $\bra{b}e^{-H}\ket{b}>0$ at
the open bonds, which restricts them to $Z^{\otimes N}\ket{b}=\ket{b}$.

\begin{figure}[t]
\includegraphics[width=\columnwidth]{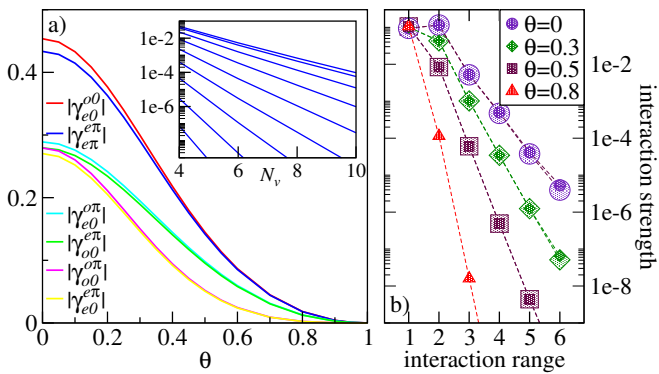}
\caption{\label{fig:rvb}
RVB--TC interpolation. \textbf{a)}
$|\gamma_{p\phi}^{p'\phi'}|$ for the off-diagonal blocks as a function of
the interpolation parameter $\theta$ (with $\theta=0$ the RVB, and
$\theta=1$ the TC), with normalization $\gamma_{e0}^{e0}=1$, for $N_v=10$.
Inset: Maximal splitting of the diagonal blocks
$|\gamma_{p\phi}^{p\phi}|$ (from top: 
$\theta=0,0.1,0.2,\dots$).
\textbf{b)} Interaction range along the interpolation for $N_v=6,8$
(small/large symbols). 
}
\end{figure}

Our findings generalize straightforwardly to cylinders with two virtual
boundaries, Fig.~\ref{fig:peps}g, as encountered when studying the ES on a
torus. From the form of the transfer operator, it is immediate that
$H'_\phi = \beta_\mathrm{topo} (P_\even\otimes P_\even+P_\odd\otimes
P_\odd) + H_\phi\otimes \openone + \openone\otimes H_\phi$, where
$\beta_\mathrm{topo}=\infty$ (i.e., the total parity must be even), and
where the non-universal $H_\phi$ is the same as before.
($H_\phi$ can differ for the two boundaries if 
$\mathbb T$ is not hermitian.)
The form of the boundary has also consequences for the topological entropy
$S(\rho_L) = S(\sigma)= \mathcal H(\{w_{p\phi}^{p\phi}\})+ \sum
w_{p\phi}^{p\phi} S(\sigma_{p\phi}^{p\phi})$, with $\mathcal H(\cdot)$ 
the Shannon entropy:
Depending on the boundary conditions,
$\mathcal H(\{w_{p\phi}^{p\phi}\})$ changes and thus the topological
correction varies between $1$ and~$-1$.
Note that 
our findings generalize to any finite group,
where the blocks of the transfer operator are labelled by the particle
types of the model~\cite{kitaev:toriccode}.

We have applied our findings to the Resonating Valence Bond (RVB) state on
the kagome lattice, and an interpolation from it to the
TC, see Appendix~D and~\cite{schuch:rvb-kagome}.  The
tensors for the RVB have a $\mathbb
Z_2$ symmetry with representation $Z=\mathrm{diag}(1,1,-1)$; additionally,
there is an $\mathrm{SU}(2)$ symmetry with representation
$\tfrac12\oplus0$. Thus, we expect the boundary Hamiltonian to describe
a system with a spinful particle or vacuum per site, with $\mathrm{SU}(2)$
invariance and conserved particle parity -- similar to a $t$--$J$ model,
but without particle number conservation.
Fig.~\ref{fig:rvb}a shows the spectrum of the transfer operator for the
RVB--TC interpolation:  The four ``diagonal'' blocks are
essentially degenerate (inset), while the off-diagonal blocks are
supressed, witnessing topological order in the system.  The boundary
Hamiltonian $H_0$ is local throughout,
see Fig.~\ref{fig:rvb}b; with dominant hopping and smaller
repulsion and Heisenberg terms at the RVB point.  These results provide
further evidence that the RVB is in the same phase as the TC, and that
its edge physics resembles a bosonic $t$--$J$ model.

\emph{Acknowledgements.---}We acknowledge helpful discussions with
F.~Verstraete.  
NS acknowledges support by the Alexander von Humboldt foundation, the
Caltech Institute for Quantum Information and Matter (an NSF Physics
Frontiers Center with support of the Gordon and Betty Moore Foundation)
and the NSF Grant No.~PHY-0803371.  DP acknowledges support by the “Agence
Nationale de la Recherche” under grant No. ANR 2010 BLANC 0406-0, and
CALMIP (Toulouse) for super-computer ressources. JIC acknowledges support
by the EU project AQUTE, the DFG SFB 631 and Exzellenzcluster NIM, and
Catalunya Caixa.  DP-G acknowledges QUEVADIS and Spanish grants QUITEMAD
and MTM2011-26912.

\appendix

\section{\label{app:boundary-state}
Entanglement spectrum for PEPS}

In this Appendix, we explain how to derive Eq.~(\ref{eq:sym-bnd}) for the
entanglement spectrum, cf.~Ref.~\cite{cirac:peps-boundaries}.  Consider a
bipartition of a PEPS into a left and a right part, and denote the
physical indices $i_1,\dots,i_{N_L}$ in the left part by $\bm
i=(i_1,\dots,i_{N_L})$, and the physical indices $i_{N_L+1},\dots,i_N$ in
the right part by $\bm j=(i_{N_L+1},\dots,i_N)$.  Further denote the
virtual indices $\alpha_1,\dots,\alpha_{N_v}$ crossing the boundary by
$\bm\alpha = (\alpha_1,\dots,\alpha_{N_v})$. Then, the total state can be
written as
\[
\ket\Psi = \sum_{\bm \alpha} L_{\bm \alpha\bm i}R_{ \bm \alpha\bm j}
\ket{\bm i}\ket{\bm j}\ ,
\]
where $L_{\bm \alpha\bm i}$ and $R_{ \bm \alpha\bm j}$ are obtained by
contracting all tensors in the left and right half, respectively. 
We define $\sigma_L = LL^\dagger$ and $\sigma_R =R R^\dagger $, as
discussed just before Eq.~(\ref{eq:sym-bnd}); moreover, we introduce the
polar decomposition of $L$, $L=PV$, where $V$ is an isometry and
$P=\sqrt{LL^\dagger} = \sigma_L$.  Then, the reduced state of $\ket\Psi$
on the left is
\begin{align*}
\rho_L &= \sum_{\bm i,\bm i'} \sum_{\bm{\alpha,\alpha',j}}
     L_{\bm \alpha\bm i}R_{ \bm \alpha\bm j} 
     L^*_{\bm \alpha'\bm i'}R^*_{ \bm \alpha'\bm j}
    \ket{\bm i}\bra{\bm i'}\\
&= L^T \, R \, R^\dagger \, L^*\\
&= V^T \, \sqrt{\sigma_L^*} \, \sigma_R \, \sqrt{\sigma_L^*} V^*\ .
\end{align*}
Since $V$ is an isometry, it follows that $\rho_L$ and $\sigma= 
\sqrt{\sigma_L^*}  \sigma_R  \sqrt{\sigma_L^*}$ have the same spectrum, 
which proves Eq.~(\ref{eq:sym-bnd}).

\section{\label{app:parity-proof}
Asymptotic equiprobability of parity blocks in the Gibbs state}

In this Appendix, we give a partial proof for the following Conjecture,
which proves that in order to obain a local boundary Hamiltonian, the
weight of the states in all sectors has to be chosen equal.

\begin{conj}
Let $H=\sum_{i=1}^N h_i$ be a Hamiltonian on a chain of $d$-level systems
of length $N$ with periodic boundaries, such that $\|h_i\|\le 1$, and each
of the $h_i$ has interaction range at most $k$. Further, there exists a
single-site operator $Z$ with eigenvalues $\pm1$ such that
$[h_i,Z^{\otimes N}]=0$ for all $i$.  Then, 
\[
\lim_{N\rightarrow\infty} 
    \frac{\mathrm{tr}[Z^{\otimes N}e^{-\beta H}]}{
	\mathrm{tr}[e^{-\beta H}]}
    = 0
\]
for any $0\le\beta<\infty$.
\end{conj}

For sufficiently small $\beta$, the conjecture can be proven using a
result of Hastings~\cite{hastings:locally}. There, it is shown that for
some $\beta<\beta^*=O(1)$ (which depends on the lattice geometry),
$\exp[-\beta H]$ can be approximated up to an error
$\epsilon=\exp[N\exp(-\ell/\xi)]-1$ in trace norm by a mixture 
\[
\rho(\ell)=
\sum_{\mathcal A} p_{\mathcal A}
    \bigotimes_{A \in \mathcal A} \rho_{A}\ ,
\]
where the sum goes over partitions $\mathcal A=(A_1,A_2,\dots)$ of the chain
into blocks $A_i$ of length at most $\ell$, the $\rho_A$ are Gibbs
states on block $A$, and $\xi$ is a constant depending on the lattice
geometry. By choosing $\ell=(1+\kappa)\xi\log N$ ($\kappa>0$), we find
that $\epsilon = \exp[N^{-\kappa}]-1\le 2 N^{-\kappa}\rightarrow0$.

On the other hand, we can bound 
\[
\left|\frac{\tr[Z^{\otimes |A|}\rho_A]}{\tr[\rho_A]}\right| \le
\frac{e^{\beta\ell}-e^{-\beta\ell}}{e^{\beta\ell}+e^{-\beta\ell}}
=\tanh(\beta\ell)\ ,
\]
and thus 
\begin{align*}
\left|\frac{\tr[Z^{\otimes N}\rho(\ell)]}{\tr[\rho(\ell)]}\right| 
&\le \tanh(\beta\ell)^{N/\ell}\\
&\le (1-\exp[-2\beta\ell])^{N/\ell}\\
&=(1-N^{-2\beta(1+\kappa)\xi})^{N/\ell}
\end{align*}
For large enough $N$, we can further bound $N/\ell\ge N^{1-\delta}$ for
any $\delta>0$, and thus (with $M=N^{1-\delta}$)
\[
\left|\frac{\tr[Z^{\otimes N}\rho(\ell)]}{\tr[\rho(\ell)]}\right| 
\le 
\left[1-M^{-\tfrac{2\beta(1+\kappa)\xi}{1-\delta}}\right]^{M}
\rightarrow 0
\]
as long as $2\beta(1+\kappa)\xi/(1-\delta)<1$, i.e.,
as long as 
\[
\beta<\tilde\beta = \frac{1}{ 2(1+\kappa)\xi}\ .
\]
Combining this with the bound on the trace norm distance between the Gibbs
state and $\rho(\ell)$, this proves the conjecture for sufficiently small
$\beta$.

\section{\label{app:ham-locality}
Analysis of the locality of the boundary Hamiltonian}

In this Appendix, we explain how we determine the locality of the
Hamiltonian by decomposing it into local terms, as shown in
Figs.~\ref{fig:ham-locality} and~\ref{fig:rvb}. 

Given a Hamiltonian $H$ on $N$ spins, we can decompose it as
\[
H=\sum_{i_1,\dots,i_N=0}^{3} c_{i_1,\dots, i_N} 
\sigma^{i_1}\otimes \sigma^{i_2}\otimes
\cdots \otimes \sigma^{i_N}\ ,
\]
where $\sigma^0, \sigma^1, \sigma^2,\sigma^3=\openone,X,Y,Z$ are
the Pauli matrices. Any term in the sum is characterized by a string
$(i_1,\dots,i_N)$; we say it has locality $d$ if the maximal distance
(with periodic boundaries) of any two $i_k\ne0$ is $d$. (Thus, $d=0$ are
one-site terms, $d=1$ are nearest neighbor terms, $d=2$ includes both
next-nearest neighbor two-body terms and true three-body terms on three
contiguous spins, etc.)

The plots show the overall weight of terms with locality $d$ as a function
of $d$, i.e., $w_d = |\vec c_d|$, where $\vec c_d$ is the vector of all
$c_{i_1,\dots,i_N}$ with locality $d$, and $|\cdot|$ is the $2$-norm.
By choosing the $2$-norm, we ensure that $w_d$ is independent of the
choice of local basis (with operator-norm normalized basis elements).
This can be seen by noting that for hermitian single-site operator, 
$O=\alpha\sigma^1+\beta \sigma^2+\gamma \sigma^3$, $\|O\|_\infty =
\sqrt{\alpha^2+\beta^2+\gamma^2}$, i.e., rotationally invariant and thus
basis-independent.

\section{\label{app:rvb}
RVB and PEPS}

For the convenience of the reader, we provide here a brief introduction to
Resonating Valence Bond (RVB) states in the PEPS formalism, as well as the
interpolation from it to the Toric Code.  We refer the reader interested
in more details to Ref.~\cite{schuch:rvb-kagome}.

\begin{figure}[b]
\includegraphics[width=\columnwidth]{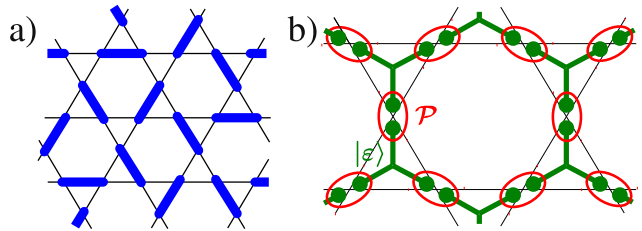}
\caption{\label{fig1}
\textbf{a)} Dimer covering of the kagome lattice.  \textbf{b)}
PEPS construction for the RVB state on the kagome lattice.
}
\end{figure}

Let us first introduce the dimer and resonating valence bond (RVB) states
on the kagome lattice.  A \emph{dimer} is a pair of vertices connected by
an edge. A \emph{dimer covering} is a complete covering of the lattice
with dimers, Fig.~\ref{fig1}a.  We can associate orthogonal quantum states
$\ket D$ with each dimer covering $D$.  Then, the \emph{dimer state} is
given by the equal weight superposition
$\ket{\Psi_\mathrm{dimer}}=\sum\ket{D}$, where the sum runs over all dimer
coverings $D$.  
Note that the dimer state is known to be locally unitarily equivalent to
Kitaev's Toric
Code~\cite{elser:rvb-arrow-representation,nayak:dimer-models,schuch:rvb-kagome}.

To obtain the RVB state, we now associate to each vertex of
the lattice a spin-$\tfrac12$ particle with basis states
$\ket{0}\equiv\lrket{\uparrow}$ and $\ket{1}\equiv\lrket{\downarrow}$.
Then, for each dimer covering $D$ we define a state $\ket{\sigma(D)}$
which is a tensor product of singlets $\ket{01}-\ket{10}$ (we omit
normalization throughout) between the pairs of spins in each dimer in the
covering ( using some consistent orientation). The \emph{resonating
valence bond} (RVB) state is then defined as the equal weight
superposition $\ket{\Psi_\mathrm{RVB}} = \sum_D \ket{\sigma(D)}$ over all
dimer coverings.

To obtain a PEPS description of the RVB state, we first place $3$-qutrit
states
\begin{equation}
       \label{eq:eps-bond}
\ket{\varepsilon} = \sum_{i,j,k=0}^2 \varepsilon_{ijk}\ket{ijk} + \ket{222}\ ,
\end{equation}
inside each triangle of the kagome lattice, as depicted in
Fig.~\ref{fig1}b. Here, $\varepsilon_{ijk}$ is the
completely antisymmetric tensor with $\varepsilon_{012}=1$, and $i$, $j$,
and $k$ are oriented clockwise.  This corresponds to having either one or
no singlet in the $\{\ket0,\ket1\}$ subspace in the triangle, the absence of a singlet being marked by
$\ket{2}$. Second, we apply the map
\begin{equation}
        \label{eq:rvb-proj}
\mc P = \ket{0}(\bra{02}+\bra{20})+\ket{1}(\bra{12}+\bra{21})
\end{equation}
at each vertex, which selects exactly one singlet per vertex.
It is straightforward to check that this construction exactly gives the
RVB state.
If we replace $\mc P$ by 
\begin{equation}
        \label{eq:orvb-proj}
\mc P_{\perp} = \ket{02}\bra{02}+\ket{12}\bra{12} + 
                     \ket{20}\bra{20}+\ket{21}\bra{21}\ ,
\end{equation}
we obtain a representation of the dimer state, since now all dimer
configurations are locally orthogonal.
Finally, we can smoothly interpolate between the dimer and the RVB (up to
isometry), by choosing
\begin{equation}
        \label{eq:P-interpolation}
\begin{aligned}
\mc P(\theta) &= 
\ket{+}\Big[\ket{0}(\bra{02}+\bra{20})+\ket{1}(\bra{12}+\bra{21})\Big]
\\
& +\theta\,\ket{-}
    \Big[\ket{0}(\bra{02}-\bra{20})+\ket{1}(\bra{12}-\bra{21})\Big]\ ,
\end{aligned}
\end{equation}
with $\theta=1$ the dimer and $\theta=0$ the RVB state, which is the
interpolation studied in Fig.~\ref{fig:rvb}.

\end{document}